\documentclass[twocolumn,showpacs,preprintnumbers,amsmath,amssymb]{revtex4}

\newcommand{\bra}[1]{\langle #1|}
\newcommand{\ket}[1]{|#1\rangle}

\newcommand{\al}[0]{\rm{Al^+}}
\newcommand{\be}[0]{\rm{Be^+}}

\newcommand{\da}[0]{|\!\!\downarrow\rangle_{Al}}
\newcommand{\ua}[0]{|\!\!\uparrow\rangle_{Al}}
\newcommand{\db}[0]{|\!\!\downarrow\rangle_{Be}}
\newcommand{\ub}[0]{|\!\!\uparrow\rangle_{Be}}

\usepackage{graphicx}
\usepackage{dcolumn}
\usepackage{bm}

\newenvironment{my_enumerate}{
\begin{enumerate}
  \setlength{\itemsep}{1pt}
  \setlength{\parskip}{0pt}
  \setlength{\parsep}{0pt}}{\end{enumerate}
}

\begin{document}

\title{High-Fidelity, Adaptive Qubit Measurements through Repetitive Information Transfer}


\author{D. B. Hume, T. Rosenband and D. J. Wineland}

\affiliation{Time and Frequency Division, National Institute of Standards and Technology, 325 Broadway, Boulder, Colorado 80305, USA}

\begin{abstract}
 Using two trapped ion species ($\rm{^{27}Al^+}$ and $\rm{^9Be^+}$)
as primary and ancillary systems, we implement qubit measurements
based on the repetitive transfer of information and quantum
nondemolition detection. The repetition provides a natural mechanism
for an adaptive measurement strategy, which leads to exponentially
lower error rates compared to using a fixed number of detection
cycles. For a single qubit we demonstrate 99.94 \% measurement
fidelity. We also demonstrate a technique for adaptively measuring
multiple qubit states using a single ancilla, and apply the
technique to spectroscopy of an optical clock transition.
\end{abstract}


\maketitle

Reliable state detection plays a central role in quantum-limited
metrology and quantum information processing. For example, in
quantum computation, low error probabilities during detection are
required to achieve good efficiency \cite{Divincenzo2001}.  In
practice, detection fidelity is limited by state perturbations and
noise during the measurement process. One way to mitigate these effects is to couple the primary quantum system to an ancillary quantum system used for measurement\cite{Divincenzo2001, Haroche2006,Schaetz2005,Schmidt2005,Gleyzes2007}. If the measurement process does not affect the projected states of the primary system, it constitutes a quantum non-demolition (QND)
measurement\cite{Caves1980,Peil1999,Meunier2006,Gleyzes2007,Lupascu2007}.
An important feature of a QND measurement is its repeatability, which allows for high fidelity state detection in the presence of noise.  The repetitive transfer of information from the primary to the ancillary system followed by detection of the ancilla state provides a natural mechanism for real-time measurement feedback, which can further enhance detection efficiency\cite{Armen2002,Cook2007}.

Ancilla-assisted detection, as formulated here, follows three  steps: (1) ancilla
preparation, (2) coupling the ancilla and primary
systems, and (3) ancilla measurement.  These steps are
described by a quantum-mechanical operator $\hat{H}(t)$, which
includes the free evolution of the states as well as the
interactions necessary for  measurement.  If this operator commutes
with the measured observable of the primary system, $\hat{O}_P$,
that is if \begin{equation} [\hat{H}(t), \hat{O}_P] = 0,\label{QND}
\end{equation} for all $t$, then it is a QND measurement\cite{Haroche2006, Caves1980}.    As an
example, consider measuring a single qubit in the
superposition state \mbox{$\alpha|\downarrow\rangle_P +
\beta|\uparrow\rangle_P$}, and let $\hat{O}_P =
\ket{\downarrow}_P\bra{\downarrow}_P$ so that $\langle \hat{O}_P
\rangle = |\alpha|^2$.  The ancilla is first prepared in a known
initial state, $|\downarrow\rangle_A$.   During subsequent
interaction with the primary system the state evolves to
\begin{equation} \alpha|\downarrow\rangle_P|\downarrow\rangle_A +
\beta|\uparrow\rangle_P|\uparrow\rangle_A.\label{Entangled1}
\end{equation}
The state of the coupled system is now detected by applying the
measurement operator for the ancillary system, $\hat{O}_A =
\ket{\downarrow}_A\bra{\downarrow}_A$. In this ideal case,
\begin{equation}
\langle \hat{O}_A \rangle = \langle \hat{O}_P \rangle.\label{APequiv}
\end{equation}
   In the presence of noisy interactions and imperfect ancilla measurements, Eq.~(\ref{APequiv}) is not strictly upheld.  But, to the degree that Eq. (1) holds and the measurement constitutes a QND measurement, it may be repeated several times to improve the measurement fidelity.  The aggregate detection after multiple cycles of information transfer and ancilla measurement projects the primary quantum system into the state $\ket{\downarrow}_p (\ket{\uparrow}_p)$ with probability $\left|\alpha\right|^2 (\left|\beta\right|^2)$.  Multiple detection cycles also allow for the measurement of multiple-qubit states by use of a single ancilla. By altering the conditions of interaction between the primary and ancillary systems the measurement can be chosen to discriminate different sets of eigenstates\cite{Haroche2006}.

\begin{figure}
\includegraphics[scale = .8]{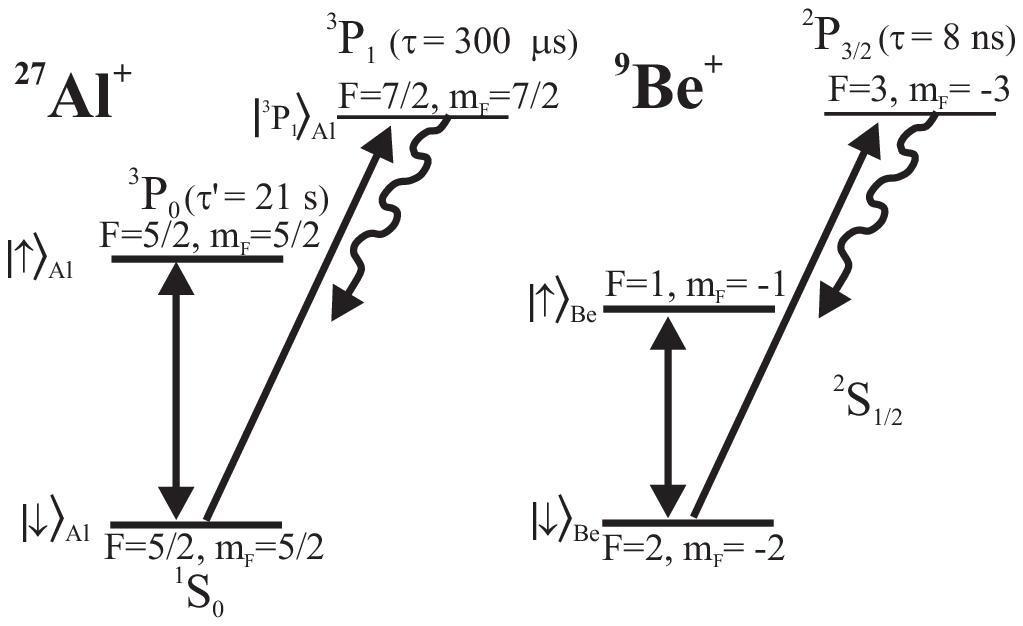}
\caption{Relevant energy levels in $\al$ and $\be$.  The states
$\da$ and $\ua$ form the qubit to be measured.  The $m_F = \frac72$
Zeeman sublevel of the $^3P_1$ state forms a closed transition with
$\da$.  The widely separated excited state lifetimes in $\al$, 21
s and 300 $\mu s$, allow for many repetitions of the detection
procedure in a single experiment. The qubit states in $\be$ are
distinguished by state-dependent resonance fluorescence\cite{Monroe1995}. }\label{fig:AtomicSystem}
\end{figure}

We apply these ideas to high-fidelity measurements of one and
two-qubit systems in an ion trap. We employ two ion
species, $\rm{^{27}Al^+}$ and $\rm{^9Be^+}$, trapped in the same
harmonic well.  The $\al$ ions form the primary quantum system and a
single $\be$ ancilla is used for sympathetic laser
cooling\cite{Barrett2003} and state detection\cite{Schmidt2005}.
Both ion species can be considered as qubits, each having an
auxillary internal state used in the detection procedure (Fig.
\ref{fig:AtomicSystem}).  In $\al$, the qubit states consist of the
\mbox{$\ket{^1S_0;F = \frac52,m_F = \frac52} \equiv\da$} ground state and
the metastable \mbox{$\ket{^3P_0;F = \frac52,m_F = \frac52} \equiv\ua$}
optically excited state. Transitions through the \mbox{$\ket{^3P_1;F =
\frac72, m_F = \frac72} \equiv\ket{^3P_1}_{Al}$} auxillary state mediate
the measurement interaction.  In $\be$, two hyperfine Zeeman
sublevels of the $^2S_{1/2}$ ground state, \mbox{$\ket{F = 2, m_F = -2}
\equiv \db$} and \mbox{$\ket{F = 1, m_F = -1} \equiv \ub$}, comprise the
qubit.  A pair of laser beams induces coherent stimulated-Raman
transitions between $\be$ qubit states. Detection of the $\be$ state
relies on state-dependent resonance fluorescence from the \mbox{$\db
\rightarrow \ket{^2P_\frac32;F = 3, m_F = -3}$} cycling
transition\cite{Monroe1995}.


Before measurement the $\al$ system is prepared in a superposition state.  The ions are then laser-cooled to the
motional ground state\cite{Barrett2003} and $\be$ is
initialized to \mbox{$\db$}\cite{Monroe1995}.  A series of laser pulses
transfers the information in the $\al$ system first to the
collective motional state then to the $\be$ internal state followed by $\be$ detection.
\noindent The individual steps and their durations are:
\begin{my_enumerate}
\item Doppler cooling all modes ($\simeq$ 600 $\mu s$)
\item Raman cooling axial modes to ground state (1 ms)
\item $\be$ preparation to the state $\db$ (1 $\mu s$)
\item Interaction between $\al$ and $\be$ ($\simeq$ 25 $\mu s$)
\item $\be$ state detection (200 $\mu s$)
\end{my_enumerate}

The procedure for generating the interaction of step 4 depends on the number of $\al$ ions to be measured and the ion configuration.  To measure the state of one $\al$ we couple the ions through the axial in-phase motional mode ($\omega_m = 2\pi\times2.62$ MHz)\cite{Schmidt2005}. Here, we denote Fock states of motion
as $|n\rangle_m$.  First, a $\pi$-pulse on
the $\da\ket{0}_m \rightarrow |^3P_1\rangle_{Al}\ket{1}_m$ sideband transition
inserts a motional quantum into the mode dependent on the ion being
in the $\da$ state. The information in the motional state is then
transferred to the internal state of $\be$ using a $\pi$-pulse on
the $\db\ket{1}_m \rightarrow \ub\ket{0}_m$ transition.  This sequence implements an entangling operation,
\begin{equation}
\big(\alpha\da + \beta\ua\big)\db \rightarrow
\alpha\ket{^3P_1}_{Al}\ub +
\beta\ua\db,
\end{equation}
leaving the system in a state analogous to that of Eq~(\ref{Entangled1}).  After measurement, the $\al$ ion is
projected into $\ua$ with probability  $|\beta|^2$.  Because the
$^1S_0$ to $^3P_1$ transition is closed, the $\al$ ion is projected into the manifold of $\da$ and $\ket{^3P_1}_{Al}$ states with probability $|\alpha|^2$. Although temporary optical excitation into $\ket{^3P_1}_{Al}$ represents a departure from the strict definition of a QND measurement, we can formally address this by defining the state $\da$ to include the eigenspace spanned by the $^1S_0$ and $^3P_1$ states.  In any case, spontaneous emission from the $^3P_1$ state ($\tau \simeq 300~\mu s$) effectively
re-prepares $\al$ back in $\da$ with probability greater than 99 \%
before another detection cycle can be implemented.

Imperfect cooling and transfer pulses give rise to a single-cycle
detection error of approximately 15 \%. However, the fidelity can be
improved by repeating the procedure. For the $j$th cycle of the
measurement procedure, a number of photons  $n_j$ is scattered from
the $\be$ ion and collected in a photomultiplier tube. The entire
measurement yields a series of photon counts, $\{n_j\}$, that are
used to determine the $\al$ state. Here, we use a computer to analyze $n_j$ in
real-time, providing the means to actively
control the measurement process.  Before the first cycle, we assume
equal prior likelihoods for $\al$ states. The probability, $P(n |
i)$, of observing $n$ photons given state $\ket{i}$ of the $\al$
system is determined based on histograms continuously updated from
previous measurements\cite{ExpFilt}. The
probability, $P(\{n_j\}|i)$, of $\ket{i}$ producing the observed
series of photon counts is $P(\{n_j\}|i)= \prod_jP(n_j | i)$.
Applying Bayes' rule,
\begin{equation}
P(i|\{n_j\}) =  \frac{P(\{n_j\}|i)}{\sum_k P(\{n_j\}|k)},
\end{equation}
yields the probability of a particular state $|i\rangle$ given the
observed series of  photon counts. Here, k spans all states
in the $\al$ system, for example the two qubit states in the case of
a single aluminum ion.  This procedure provides both the most likely
state of $\al$, $|i_{max}\rangle$, and also the probability of
measurement error, $1-P(i_{max}|\{n_j\})$. The first assigns
detection outcomes, while the second is used to optimize the
measurement process. Specifically, we
repeat the detection cycles only until the aggregate detection
reaches a desired error probability.

\begin{figure}
\includegraphics[scale = 0.4]{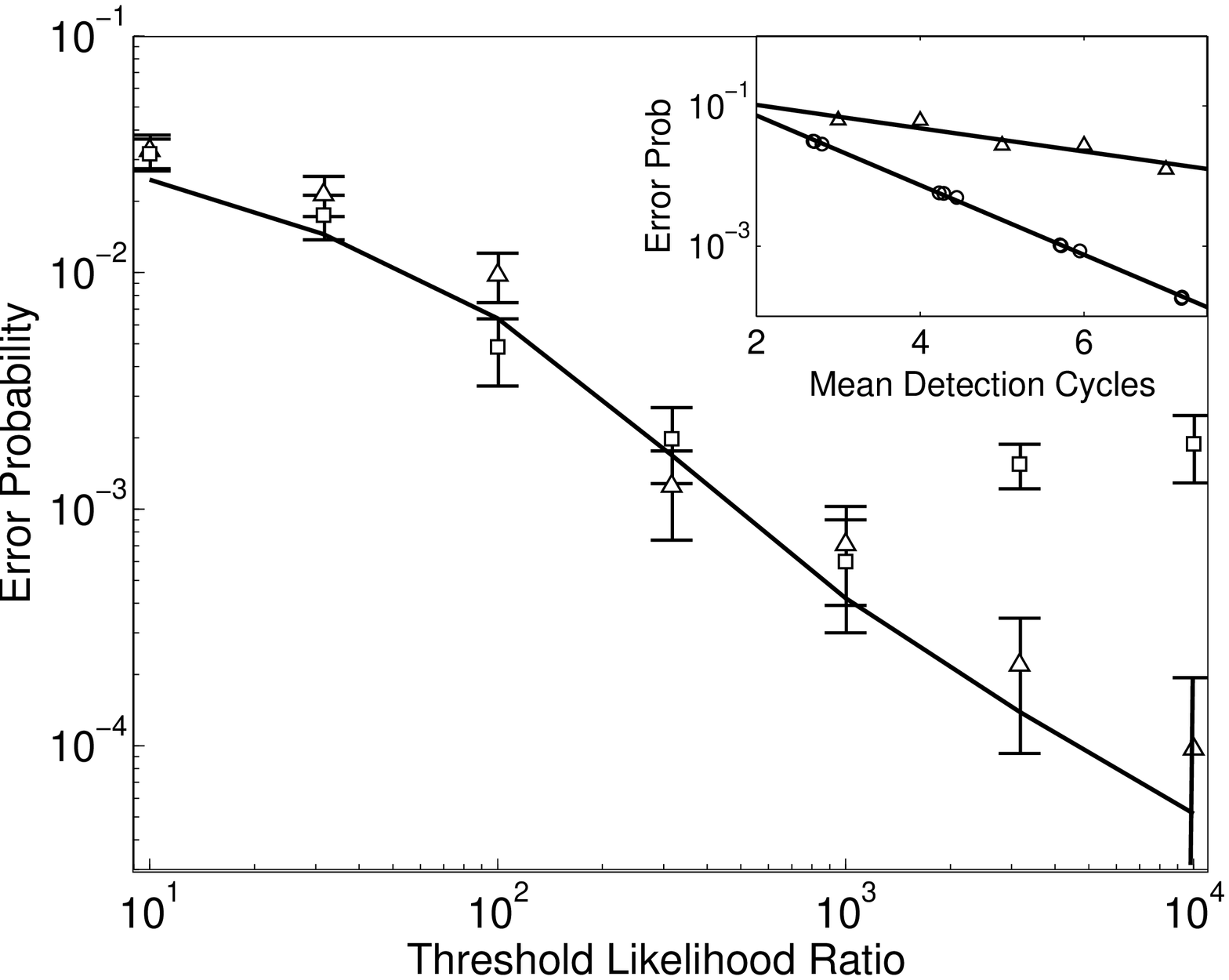}
\caption{Observed (symbols) and calculated (solid line) error rates
for repeated detections are plotted as a function of the threshold
likelihood ratio defined as the greater of
\mbox{$P(\{n_j\}|\ua)/P(\{n_j\}|\da)$} or
\mbox{$P(\{n_j\}|\da)/P(\{n_j\}|\ua)$}. Once the likelihood
ratio exceeds the threshold value, detection is
stopped.  The $^1S_0$ detections (triangles) reach a repeatability
of 99.99 \% while $^3P_0$ detections (squares) are limited by the
state lifetime to 99.94 \%.  This is achieved for a desired likelihood
ratio of $10^3$, requiring a mean number of detection cycles equal
to 6.54. Inset: Simulation of qubit detection, using experimental
histograms, comparing the case in which the number of detection
cycles is fixed(diamonds) to that in which it is adaptive(circles).The ability to estimate errors in real time gives
rise to an exponentially lower error rate as a function of
time.}\label{fig:DetData}
\end{figure}

To experimentally determine the error rate for state discrimination,
we compare two consecutive detection sequences, each of which separately determines $\ket{i}$ and reaches a specified minimum
error probability. If the two results agree, both detections are
counted as correct, while disagreement signifies an error. This analysis
allows us to compare the actual error rate with the real-time
prediction.   Errors
quantified in this way also represent the fidelity of quantum state
preparation\cite{Lupascu2006}. The results for a single $\al$ ion are plotted in
Fig.~\ref{fig:DetData}. The observed errors agree well with the
predicted error rate for fidelities up to 99.94 \%.  In the case of
detecting $\da$, the observed error rate is as low as
$9\times10^{-5}$. However, $\ua$ detections, the state
lifetime of 21 s limits the number of times the detection cycle can be
repeated and still yield an accurate prediction. Here, the observed
error rate reaches a minimum at $6\times10^{-4}$, then increases to
above $1\times10^{-3}$ as we demand higher measurement confidence
through more repetitions. This error rate agrees with that predicted from the decay rate and the interval between detection cycles. The $\ua$ state lifetime $\tau '$ and
measurement cycle duration, $t_c$ set an upper bound on the
attainable detection fidelity; that is, no measurement can achieve
an error rate lower than the probability of decay before the first
detection, $t_c/\tau ' \simeq 10^{-4}$ in the experiment
here.

To demonstrate the gain in sensitivity achieved with adaptive measurements, we perform a Monte-Carlo simulation of the
detection procedure based on experimentally observed histograms. We
compare the adaptive scheme, which uses the minimum number of cycles
necessary to achieve a given fidelity, to a detection scheme
where the number of detection cycles is set to a particular value
[Fig.~\ref{fig:DetData}(inset)]. As a function of
average detection duration, the adaptive detection gives an
exponentially smaller error rate.  In both simulations the final state determination results from Bayesian analysis, which gives optimal results based on the detection record.  Thus, in the case of fixed detections, the plotted error rate provides a lower bound for the  infidelity of commonly used fixed detection strategies such as majority vote.

\begin{figure}
\includegraphics[scale = .4]{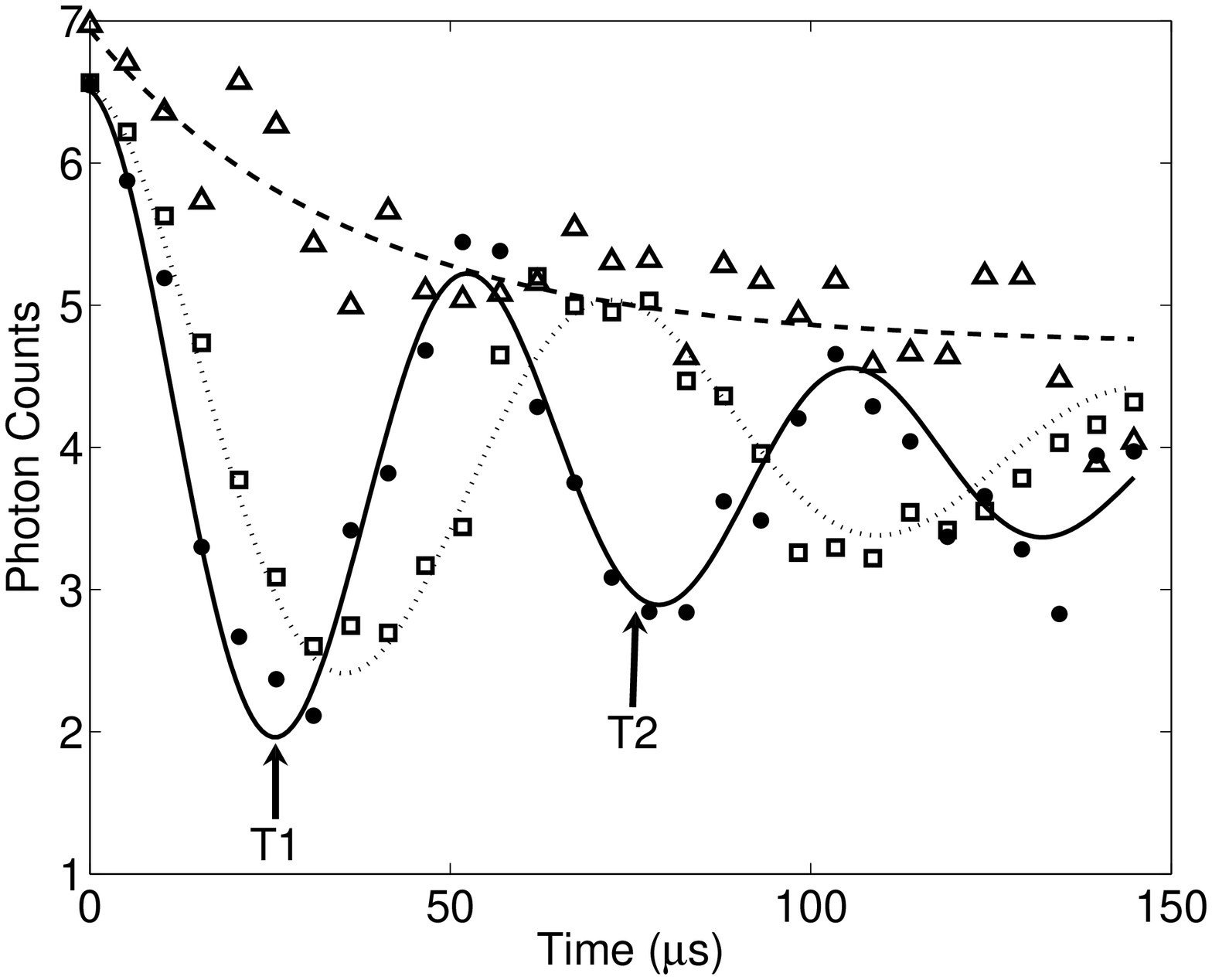}
\caption{Red sideband Rabi flopping on the $\da - |^3P_1
\rangle_{Al}$ transition beginning with one quantum of motion. The
fluorescence signal is obtained after transferring the motional
information to the qubit state of $\be$. The three possible numbers
of aluminum ions in the ground state, 2 (solid circles), 1 (empty
squares), and 0 (empty triangles), are distinguished by their
sideband Rabi rate.  For zero ground state ions, the known signal admixture from one ground state ion due to detection errors was removed.  We determine the state of excitation in the aluminum ion
system after making multiple mapping sequences with pulses of
duration T1 or T2.}\label{fig:SidebandRabi} \label{fig:3P1Flop}
\end{figure}

Adaptive detection also provides a means to measure and prepare the state of a multiple qubit system without the need for individual qubit addressing.  Although a single ancilla qubit can yield at best one bit of information in a measurement cycle, repeating the measurement process and varying the parameters of the interaction yields more information\cite{Haroche2006}.  To demonstrate the detection of two $\al$ ions, we use a symmetric spatial
configuration of the ions ($\be$ centered between the two $\al$
ions).  First, all nine normal modes are Doppler cooled. The antisymmetric modes, which normally don't couple to $\be$, are Doppler cooled sympathetically by first distorting the ion configuration with application of a static field and then adiabatically relaxing the system back to the aligned
configuration.  The two axial modes that couple to $\be$ are cooled to the ground state and $\be$ is prepared in $\db$.

  The measurement interaction for two $\al$ ions proceeds as follows:
\begin{my_enumerate}
\item $\be$: $\db|0\rangle_m \rightarrow \ub|1\rangle_m$ ($T_\pi$).
\item $\al$: $\da|1\rangle_m \rightarrow \ket{^3P_1}_{Al}|0\rangle_m$ ($T_{var}$).
\item $\be$: $\ub|1\rangle_m \rightarrow \db|0\rangle_m$ ($T_\pi$).
\end{my_enumerate}
where $T_\pi$ is the sideband $\pi$-time for the $\be$ qubit
transition and $T_{var}$ is a variable duration.  The first pulse inserts a quantum of motion
into the selected motional mode, the second pulse entangles the
motional state with the internal $\al$ qubit pair, and the final
pulse transfers information in the motional state to the internal
state of $\be$.

The $^3P_1$ sideband Rabi rate (step 2) depends only on the number of $\da$ ions (zero, one or two).  Experimental excitation curves for these three cases are plotted in Fig.~\ref{fig:3P1Flop}. With one $\al$ ion in $\da$, flopping between $\da\ket{1}_m$ and $\ket{^3P_1}\ket{0_m}$ will proceed with a Rabi rate given by $\Omega_{R,1} \simeq \Omega_c\eta$, where $\Omega_c$ is the carrier Rabi frequency and $\eta$ is the Lamb-Dicke parameter for the aluminum ions\cite{Monroe1995}.  With two $\al$ ions in $|\downarrow \rangle_{Al}$, Rabi flopping carries the aluminum system to an entangled state, $\da\da\ket{1}_m
\boldsymbol{\rightarrow}\frac1{\sqrt{2}} \big(\ket{^3P_1}\da +
\da\ket{^3P_1}\big)\ket{0}_m$ with characteristic Rabi rate
$\Omega_{R,2} \simeq \sqrt{2}\Omega_c\eta$\cite{King1998}.  The fitted Rabi
rates agree with those expected.

\begin{figure}
\includegraphics[scale = .34]{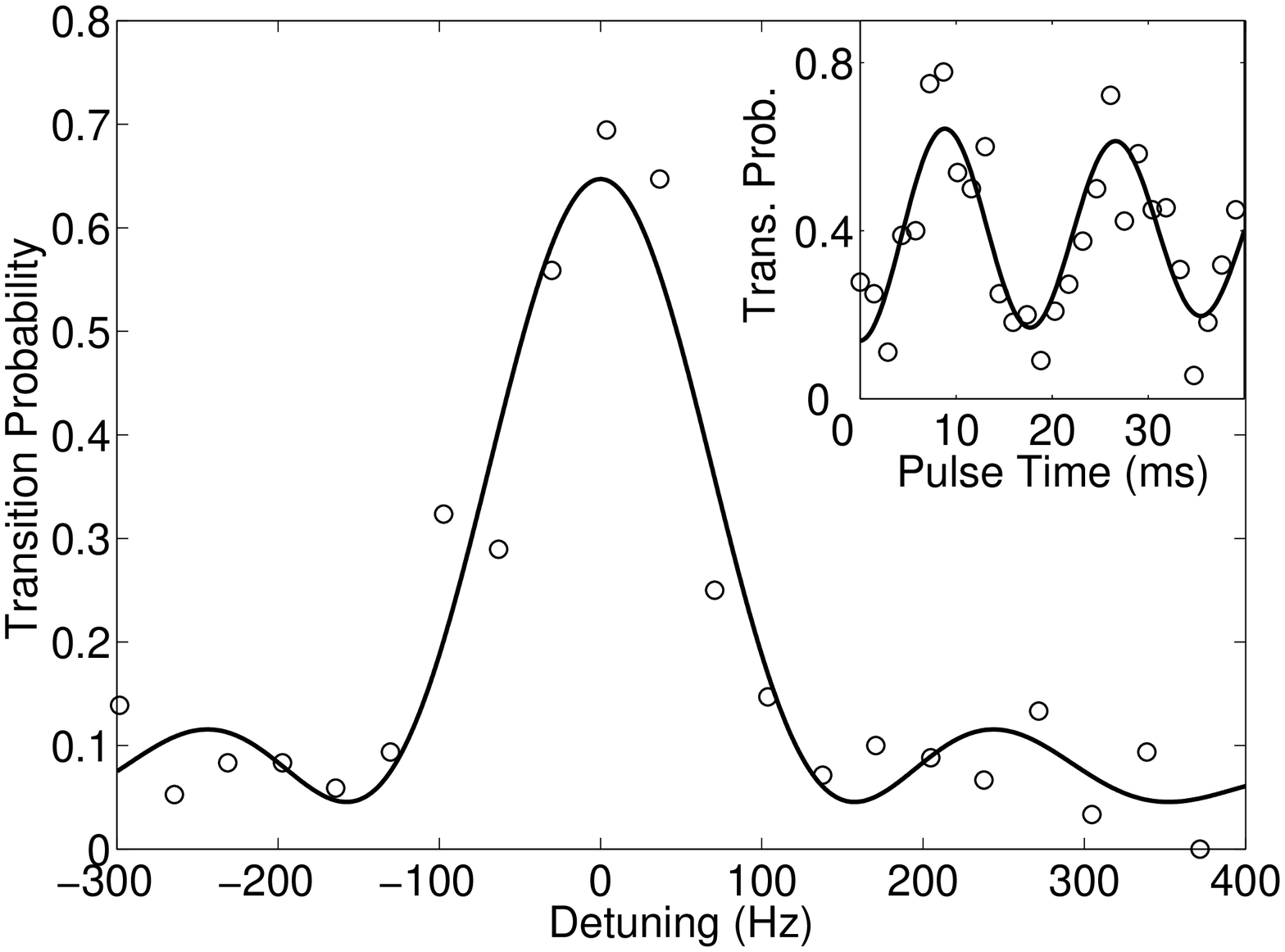}
\caption{Signal from $^3P_0$ spectroscopy using two $\al$ and one $\be$ ancilla.
Obtaining this signal depends on the ability to prepare the $\da
\da$ or $\ua \ua$ state of the $\al$ system because transitions
between $\da \ua$ and $\ua \da$ will not be detected.  Inset:  Rabi
flopping on the $^3P_0$ transition.  Signal contrast is limited by
fluctuations in the transition excitation, rather than detection
efficiency.}\label{fig:3P0Sig}
\end{figure}

Two particular transfer pulse durations, \mbox{$\rm{T1} = 30~ \mu s$} and \mbox{$\rm{T2} = 80~ \mu s$}, exhibit good discrimination between one particular state and the other two.  A
combination of detection cycles using these two pulse durations
distinguishes the three states.  In this scheme, the maximum
likelihood analysis plays the additional role of determining which
pulse duration to use based on previous measurements. The duration is chosen to maximize contrast between the two most-likely states of the aluminum system.

We have performed spectroscopy  of the $\da \rightarrow \ua$
transition on two $\al$ ions using this scheme.
Figure~\ref{fig:3P0Sig} shows  a Fourier-limited lineshape and Rabi
flopping (inset). Various sources of noise including ion
temperature, laser intensity noise and off-resonant
excitation of the ground state into the $^3P_1, F = \frac72, m_F =
\frac52$ state due to imperfect polarization are the primary limits to signal contrast.  As before, we measure the error rate for state discrimination by comparing consecutive detection sequences, and find a detection fidelity of $98.3\%$.  Improvements in laser cooling could improve signal contrast and detection fidelity.


A practical application for the techniques we describe is the
operation of an optical atomic clock based on
$\al$ \cite{Rosenband2007}.  The time scale for probing the \mbox{$^1S_0 \rightarrow ^3\!\!P_0$} transition (100 ms) significantly exceeds the detection duration ($\simeq$ 10
ms) so that the clock performance is not significantly affected by
the detection duration. These techniques may also be important for scalable quantum computation.  Efficient state detection is a basic requirement, which may ultimately depend on repeated QND measurements as demonstrated here.  More specifically, trapped ion systems composed of two species have previously been proposed for large-scale QIP \cite{Wineland1998,Kielpinski2002}.  Here, one species carries the qubit, and the other provides sympathetic cooling. Such a two-species system could also utilize an analogous protocol to the one we describe, thereby reaching very high detection fidelity in a minimal time period.

This work is supported by ONR and DTO.  We thank R. J. Epstein, J. J. Bollinger and E. Knill for helpful comments on the manuscript.  Contribution of NIST; not subject to U. S. copyright.


\begin{thebibliography}{18}
\bibitem[{\citenamefont{ExpFilt}(2003)}]{ExpFilt}
\bibnamefont{Updates are made by applying an exponential
filter to previous measurement results, providing accurate
histograms even if experimental parameters slowly drift.}
\end{thebibliography}

\end{document}